\documentstyle[12pt,epsfig]{article}
\makeatletter
\def\slash#1{{\mathpalette\c@ncel{#1}}} 
\makeatother

\newcommand\beq{\begin{eqnarray}}
\newcommand\eeq{\end{eqnarray}}

\def\xhat{\widehat{x}}
\def\zhat{\widehat{z}}

\def\pvec{\vec{p}}

\begin{document}
\begin{flushright}
BNL-NT-06/9 \\
RBRC-586 \\  
\today
\end{flushright}
\vspace*{15mm}
\begin{center}
{\Large \bf Resummation for Polarized Semi-Inclusive \\[4mm]
Deep-Inelastic Scattering at Small \\[6mm] Transverse Momentum}
\vspace{1.5cm}\\
 {\sc Yuji Koike$^1$, Junji Nagashima$^1$, Werner Vogelsang$^2$}
\\[0.4cm]
\vspace*{0.1cm}{\it $^1$ Department of Physics, Niigata University,
Ikarashi, Niigata 950-2181, Japan}\\
\vspace*{0.1cm}{\it $^2$ Physics Department and RIKEN BNL Research Center, 
Brookhaven National Laboratory, Upton,
NY 11973, USA}
\\[3cm]

{\large \bf Abstract} \end{center}

\noindent
We study the transverse-momentum distribution of hadrons
produced in semi-inclusive deep-inelastic scattering (SIDIS). We
consider cross sections for various combinations of
polarizations of the initial lepton and nucleon or the produced
hadron, for which we perform the resummation of large double-logarithmic
perturbative corrections arising at small transverse momentum. We present
phenomenological results for the processes $lp\to l\pi X$ with 
longitudinally polarized leptons and protons. We discuss the 
impact of the perturbative resummation and of estimated non-perturbative 
contributions on the corresponding cross sections and their spin 
asymmetry. Our results should be
relevant for ongoing studies in the COMPASS experiment at CERN, and
for future experiments at the proposed eRHIC collider at BNL.

\newpage

\section{Introduction}

Our knowledge about the structure of hadrons has been vastly improved
by experiments with polarized high-energy lepton beams scattering
off polarized nucleon targets. Spin observables in deeply-inelastic 
lepton-nucleon collisions allow us to extract the spin-dependent parton 
distributions of the nucleon. At the same time, they challenge our
understanding of the reaction mechanism within QCD, and our
ability to perform reliable theoretical calculations of the relevant
cross sections for each process and kinematic region of interest.

Of particular interest in this respect are observables in semi-inclusive 
deeply-inelastic scattering (SIDIS), $lp\to lhX$, for which a hadron $h$ is 
detected in the final state. Depending on the type of hadron
considered, various different aspects of nucleon structure may be probed.
Two current lepton scattering experiments, HERMES at DESY and COMPASS at CERN,
employ this method extensively. Measurements of spin asymmetries for
longitudinally polarized beam and target, integrated over all transverse
momenta of the produced hadron, have served to allow conclusions about
the helicity-dependent up, down, and strange quark and anti-quark 
distributions in the nucleon~\cite{sidisdata}. It has also been 
recognized that distributions in the hadron's transverse momentum can be
of great interest, in particular when the nucleon target is transversely 
polarized~\cite{trrev}. The associated experimental investigations by 
HERMES~\cite{hermes}, COMPASS~\cite{compass}, the SMC~\cite{smc}, and 
CLAS~\cite{jlab} have been remarkably productive and have opened
windows on novel QCD phenomena such as the Sivers~\cite{sivers} 
and Collins~\cite{collins} effects. It is hoped that experiments
at a possibly forthcoming polarized electron-proton collider, eRHIC,
would carry on and extend these studies~\cite{erhic}.

Theoretically, the most interesting kinematic regime is characterized
by large virtuality $Q^2$ of the photon exchanged in the DIS process,
and relatively small transverse momentum, $q_T\ll Q$. This regime also
provides for the bulk of the events in experiment. 
It is precisely here, for example, that effects related to 
intrinsic transverse momenta of partons in the nucleon may become visible,
potentially offering new insights into nucleon structure. At the same
time, the theoretical analysis of hard-scattering in this regime is
fairly involved, but well-understood. In particular, the emission of
gluons from the DIS Born process $\gamma^{\ast}q\to q$
also leads to non-vanishing transverse-momentum of the final-state
hadron and needs to be taken into account appropriately. It is the goal of this 
paper to present state-of-the-art calculations for the transverse-momentum 
dependence of some SIDIS observables. For this study, we will focus
entirely on the set of leading-twist {\it double-spin} reactions,
\beq
&{\rm (i)}&\quad e+p\rightarrow
e+\pi+X \; ,\nonumber\\
&{\rm (ii)}&\quad e+\vec{p}\rightarrow e+ \vec{\Lambda}+X \; ,\nonumber\\
&{\rm (iii)}&
\quad e+ {p}^\uparrow\rightarrow e+ {\Lambda}^\uparrow +X \; ,\nonumber\\
&{\rm (iv)}&
\quad \vec{e}+ \vec{p}\rightarrow e+ \pi+X \; ,
\nonumber\\
&{\rm (v)}&\quad \vec{e}+ p\rightarrow e+ \vec{\Lambda} +X \; . 
\label{eq1.1}
\eeq
Here arrows to the right (upward arrows) denote longitudinal (transverse)
polarization. Needless to say that the final-state pion could be replaced
by any hadron. The same is true for the $\Lambda$, as long as the observed 
hadron is spin-1/2 and its polarization can be detected experimentally.
In our study, we will make use of several
ingredients available in the literature. In an earlier publication~\cite{KN03}, 
two of us presented results for the $2\to 3$ partonic reactions $lq
\to lqg$ and $lg\to lq\bar{q}$, for all polarizations of interest. In the
perturbative expansion and using collinear factorization, these processes are 
the first to yield a non-vanishing transverse momentum of the produced hadron. 
In terms of the strong coupling $\alpha_s$ they are of order ${\cal O}(\alpha_s)$.
We therefore refer to them as ``leading order (LO)'' processes for the 
hadron transverse-momentum distribution. They are expected to be adequate 
(at least qualitatively) for achieving a good theoretical description at large 
transverse momentum, $q_T\sim Q$. In the unpolarized case, the complete 
next-to-leading (NLO) (${\cal O}(\alpha_s^2)$) corrections to the 
$q_T$-distribution have been calculated~\cite{Aur,NLOqT,Kn} which will lead to an 
improvement
of the theoretical calculation at large $q_T$.~\footnote{The 
study~\cite{KN03} may be viewed as an extension of 
previous work on the $q_T$-{\it integrated} polarized SIDIS cross section, for 
which the LO process is $\gamma^{\ast}q\to q$. LO calculations were performed
in~\cite{Ji94} and NLO ones in~\cite{dGS,dSV98} (for initial work on the NNLO 
corrections to the unpolarized $q_T$-integrated SIDIS cross section, 
see~\cite{NNLOSIDIS}).}

At low transverse momentum, $q_T\ll Q$, fixed-order calculations are bound to fail.
The reason for this is well understood: when $q_T\to 0$, gluon
emission is inhibited, and the cancellation of infra-red
singularities between real and virtual diagrams in the perturbative
series leaves behind logarithmic remainders of the form
\beq
\alpha_S^k\,\frac{\ln^{m}\left(Q^2 /q_T^2\right)}{q_T^2} \label{eq1.2}
\eeq
in the cross section $d\sigma/dq_T^2$
at the $k$th order of perturbation theory, where $m=1,\ldots, 2k-1$. Ultimately,
when $q_T\ll Q$, $\alpha_s$ will not be useful anymore as the expansion 
parameter in the perturbative series since the logarithms will compensate
for the smallness of $\alpha_s$. Accordingly, in order to obtain a 
reliable estimate for the cross section, one has to sum up (``resum'') 
the large logarithmic contributions to all orders in $\alpha_s$.
Techniques for this resummation are well established, starting
with pioneering work mostly on the Drell-Yan process in the late 
1970's to mid 1980's~\cite{DDT78,PP79,CS81,AEGM84,CSS85}. The 
``Collins-Soper-Sterman'' (CSS) formalism~\cite{CSS85} has become the 
standard method for $q_T$ resummation. It is formulated in 
impact-parameter ($b$) space, which guarantees conservation of the soft-gluon 
transverse momenta. The formalism has also been applied to the unpolarized 
SIDIS cross section~\cite{MOS96,NSY00}, and it was found~\cite{NSY00} 
that data for $q_T$ distributions from the HERA $ep$ collider~\cite{h1,zeus} 
are satisfactorily described.

Resummations at small transverse momentum $q_T$ have also been 
developed for spin observables. In Refs.~\cite{weber,att} the CSS
formalism was applied to longitudinal and transverse double-spin asymmetries 
in the Drell-Yan process. Early resummation studies on $q_T$-distributions 
in jet production in polarized DIS were performed in~\cite{weberthesis}.
In Ref.~\cite{dboer,dboer1} leading-logarithmic (LL) resummation 
effects were investigated for spin asymmetries that involve 
transverse-momentum dependent distributions, in particular the 
Collins functions mentioned above. In Ref.~\cite{yuan}, resummation 
formulas for polarized SIDIS were derived, based on 
a factorization theorem at low transverse momentum~\cite{yuan1}.
A main result of~\cite{yuan} is that the CSS evolution equation is
the same for the spin-dependent cases as in the unpolarized one.
Knowing the expression for the unpolarized resummed cross section, 
and the complete polarized LO cross sections~\cite{KN03}, it is then relatively
straightforward to determine the resummed expressions for the polarized
case. These will be provided in explicit form in this paper, for
the cases listed in~(\ref{eq1.1}). We shall present results corresponding
to resummation to next-to-leading logarithmic (NLL) accuracy, which 
corresponds to resummation of the towers with $2k-1,2k-2,2k-3$ in
(\ref{eq1.2}). We shall also present numerical 
estimates for the reaction $\vec{e}\vec{p}\to e\pi X$ at COMPASS
and eRHIC, in order to study the general  features of the resummation
and its impact on the cross sections and the spin asymmetry. 
We note that in the numerical evaluation one needs to specify a 
recipe for treating the integration over the impact parameter at very
large $b\sim 1/\Lambda_{\rm QCD}$, in order to 
avoid the Landau pole present in the resummed expression. 
This is closely related to non-perturbative effects generated by 
resummation~\cite{CSS85,DS84,DYbstar,Kony,KSV02,Qiu,Tafat}. 
We will use two different methods for treating the large $b$-region.

We stress that we will not address single-transverse spin phenomena in this
paper like those related to the Sivers and Collins effects mentioned
above. For these, the analysis of QCD radiative effects is rather
more involved~\cite{dboer1} than for the double-spin case we consider, 
in particular regarding the connection of the behavior at small and 
large transverse momentum $q_T$~\cite{prep1}. Also, as we shall see 
below, the $q_T$-differential
cross section in general contains terms depending on the angle $\phi$
between the hadron and lepton planes. We shall only consider the
resummation of the $\phi$-independent pieces in the cross section,
which dominate at small $q_T$. It is an interesting topic by itself to
study the resummation of the terms that depend on $\phi$~\cite{prep2}.

The remainder of this paper is organized as follows: Section~\ref{sec2}
provides all formulas needed for the $q_T$-differential SIDIS cross
section, at LO and for the NLL resummed case. In section 3, 
we present numerical estimates for $\vec{e}\vec{p}\to e\pi X$ at COMPASS
and eRHIC, based on the resummation formulas. We discuss in particular 
the effect of the resummation on the spin asymmetry. We also study the
impact of the treatment of the large-$b$ region and of non-perturbative 
corrections on the resummed cross sections and the spin asymmetry. We
conclude in Sec.~\ref{sec4}. In the Appendix, we list the complete $O(\alpha_s)$ 
cross sections for the processes in (\ref{eq1.1}), correcting some typos 
in~\cite{KN03}. 

\section{The $q_T$-differential SIDIS cross section \label{sec2}}

\subsection{Kinematics \label{sec2.1}}

We are interested in the cross section for the process
\beq
\vec{l}(k)+A(p_A,S_A) \rightarrow l(k') + B(p_B,S_B)+X \; ,
\label{eq2.1}
\eeq
where $S_A,S_B$ are the spin vectors for the initial nucleon
and the produced final-state hadron, which can be longitudinal
or transverse, as indicated in~(\ref{eq1.1}). The lepton can
be unpolarized or longitudinally polarized; transverse-spin effects
for the lepton are suppressed by $m_l/Q$ because of chirality conservation
at the lepton-photon vertex and hence are negligible. From now
on, we will for definiteness take hadron $A$ to be a proton and the 
lepton $l$ to be an electron. We define five Lorentz invariants, denoted
$S_{ep}, x_{bj},Q^2,z_f$, and $q_T^2$, to describe 
the process. The center of mass energy squared, $S_{ep}$, for the initial
electron and the proton is
\beq
S_{ep}=(p_A + k)^2 \simeq 2p_A\cdot k \; ,
\eeq
ignoring masses. The conventional DIS variables are defined 
in terms of the virtual photon momentum $q=k-k'$ as
\beq
x_{bj}={Q^2\over 2p_A\cdot q} \; , \qquad Q^2 =-q^2=-(k-k')^2 \; ,
\eeq
and they may be determined experimentally by observing the scattered 
electron. For the final-state hadron $B$, we introduce the scaling variable
\beq
z_f={p_A\cdot p_B\over p_A\cdot q} \; .
\eeq
Finally, we define the ``transverse'' component of $q$, which is orthogonal to
both $p_A$ and $p_B$:
\beq
q_t^\mu=q^\mu- {p_B\cdot q\over p_A\cdot p_B}p_A^\mu -
{p_A\cdot q\over p_A\cdot p_B}p_B^\mu \; .
\eeq
$q_t$ is a space-like vector, and we denote its magnitude by
\beq
q_T = \sqrt{-q_t^2} \; .
\eeq
To completely specify the kinematics, we need to choose a reference
frame. We shall work in the so-called {\it hadron frame}~\cite{MOS92,Nad},
which is the Breit frame of the virtual photon and the initial proton:
\beq
q^\mu &=& (0,0,0,-Q) \; ,\\
p_A^\mu &=& \left( {Q\over 2x_{bj}},0,0,{Q\over 2x_{bj}}\right) \; .
\eeq
Further, in this frame the outgoing hadron $B$ is taken to be in the $xz$ plane:
\beq
p_B^\mu = {z_f Q \over 2}\left( 1 + {q_T^2\over Q^2},{2 q_T\over Q},0,
{q_T^2\over Q^2}-1\right) \; .
\label{eq2.p_B}
\eeq
As one can see, the transverse momentum of hadron $B$ is in this 
frame given by $z_f q_T$. This is true for any frame in which the
3-momenta of the virtual photon and the initial proton are collinear.
By introducing the angle $\phi$ between the hadron plane and the lepton plane, 
the lepton momentum can be parameterized as
\beq
k^\mu={Q\over 2}\left( \cosh\psi,\sinh\psi\cos\phi,
\sinh\psi\sin\phi,-1\right) \; ,
\label{eq2.lepton}
\eeq
and one finds
\beq
\cosh\psi = {2x_{bj}S_{eA}\over Q^2} -1 \; .
\label{eq2.cosh}
\eeq
For the case (iii) in~(\ref{eq1.1}) [$ep^\uparrow\to e\Lambda^\uparrow X$] 
when the initial proton is transversely polarized and the transverse polarization 
of an outgoing spin-1/2 hadron is observed, we need to parameterize the
spin vectors $S_{A\perp}$ and $S_{B\perp}$ to lie in the planes orthogonal to 
$\pvec_A$ and $\pvec_B$, respectively. In the hadron frame, 
they can be written as
\beq
S_{A\perp}^\mu &=&(0,\cos\Phi_A,\sin\Phi_A,0) \; ,\nonumber\\
S_{B\perp}^\mu &=&(0,\cos\Theta_B\cos\Phi_B,\sin\Phi_B,-\sin\Theta_B\cos\Phi_B) \; ,
\eeq
so that $\Phi_{A,B}$ are the azimuthal angles of $S_{A,B\perp}$ around
$\vec{p}_{A,B}$
as measured 
from the hadron plane 
and $\Theta_B$ is the polar angle of $\pvec_B$ 
measured with respect to $\pvec_A$. One finds:
\beq
\cos\Theta_B={q_T^2-Q^2\over q_T^2+Q^2},\qquad \sin\Theta_B=
{2q_T Q\over q_T^2+Q^2} \; .
\eeq
We note that the polarization state depends on the frame we choose.  
For example, a state that is transversely polarized in the
hadron frame becomes a mixture of longitudinally polarized and
transversely polarized states in the laboratory frame where the 
initial electron and proton are collinear. With the above definitions, 
the cross section for (\ref{eq2.1}) can be expressed in terms of
$S_{ep}$, $x_{bj}$, $Q^2$, $z_f$, $q_T^2$ and $\phi$ in the hadron
frame. Note that $\phi$ is invariant under boosts in the 
$\vec{q}$-direction, so that is the same in the hadron frame and, 
for example, in the photon-proton center-of-mass frame.

\subsection{Structure of the lowest-order cross section \label{sec2.2}}

The complete set of the $O(\alpha_s)$ spin-dependent partonic cross 
sections for large-$q_T$ hadron production in SIDIS have been derived in 
Ref.~\cite{KN03}. They are obtained from the Feynman diagrams for the reactions
$lq\to lqg$ and $lg\to lq\bar{q}$. For completeness, and for the reader's
convenience, we list them in the Appendix. Here we summarize the main 
characteristics of the hadronic cross section.  

The cross section can be decomposed into several pieces with 
different $\phi$-dependences:
\beq
{d^5\sigma\over dQ^2dx_{bj}dz_fdq_T^2 d\phi}
=\sigma_0 + {\rm cos}(\phi)\sigma_1 +
{\rm cos}(2\phi)\sigma_2 \; ,
\label{eq3unpol}
\eeq 
for processes (i) and (ii) in~(\ref{eq1.1}), 
\beq
{d^5\sigma\over dQ^2dx_{bj}dz_fdq_T^2 d\phi}
=\sigma_0 + {\rm cos}(\phi)\sigma_1 \; ,
\label{eq3pol}
\eeq 
for (iv) and (v), and 
\beq
\hspace*{-6mm}
{d^5\sigma_T\over dQ^2dx_{bj}dz_fdq_T^2 d\phi}
=\cos(\Phi_A-\Phi_B-2\phi)\sigma_0^T + \cos(\Phi_A-\Phi_B-\phi)\sigma_1^T
+\cos(\Phi_A-\Phi_B)\sigma_2^T \; ,  \label{eq3Tpol}
\eeq 
for (iii). In the hadron frame, $\sigma_0$ and $\sigma_0^T$ in these 
decompositions behave as $\alpha_s \ln\left(Q^2/q_T^2\right)/q_T^2$
at small $q_T$. This is the manifestation of the large logarithmic
corrections described in the Introduction at this order. At yet higher
orders, corrections as large as $\alpha_S^k\,\ln^{2k}\left(Q^2 /q_T^2\right)/
q_T^2$ in the cross section arise. In the following, we will study the NLL 
resummation of these large logarithmic corrections within the CSS formalism. 
We note that the other components of the cross sections 
in Eqs.~(\ref{eq3unpol})-(\ref{eq3Tpol}) are less singular as $q_T\to 0$:
$\sigma_1$ and $\sigma_1^T$ behave as $\alpha_s \ln\left(Q^2/q_T^2\right)/q_T$, 
and $\sigma_2$ and $\sigma_2^T$ as $\alpha_s \ln\left(Q^2/q_T^2\right)$. 
These pieces too, however, receive large higher-order corrections at small
$q_T$~\cite{MOS96} and would need to be resummed~\cite{prep2}. In this paper 
we will only focus on the resummation for $\sigma_0$ and $\sigma_0^T$.

\subsection{Asymptotic part of the lowest-order cross section \label{sec2.3}}

We will now discuss the behavior of the cross sections
at small $q_T$ in more detail. This is a first step in arriving at the 
resummed cross section. Following \cite{MOS96,NSY00}, we write the cross section
at $O(\alpha_s)$ as
\beq
{d^5\sigma^{\rm LO}\over dQ^2dx_{bj}dz_fdq_T^2 d\phi}
={d^5\sigma^{\rm asymp}\over dQ^2dx_{bj}dz_fdq_T^2 d\phi}
+Y(S_{ep},Q,q_T,x_{bj},z_f) \; ,
\label{asymptotic}
\eeq
where the first term on the right-hand-side is the ``asymptotic''
part in $\sigma_0$ and $\sigma_0^T$ containing the pieces that are singular as 
$\ln\left(Q^2/q_T^2\right)/q_T^2$ or $1/q_T^2$ at small $q_T$, 
while $Y$ collects the remainder which is at most logarithmic as $q_T\to 0$. 
It is straightforward to derive the asymptotic part of the LO unpolarized 
SIDIS cross section from the expressions in the Appendix. One finds:
\beq
{d^5\sigma^{\rm asymp}\over dQ^2dx_{bj}dz_fdq_T^2 d\phi}
&=&{\alpha_{em}^2 \alpha_s \over 8\pi x_{bj}^2 S_{ep}^2 Q^2} \,
{\cal A}_1\, \frac{2Q^2}{q_T^2}\,\sum_{q,\bar{q}}\,e_q^2
\left[
2f_q(x_{bj},\mu) D_q(z_f,\mu)\left(C_F\ln\left({Q^2\over q_T^2}\right) -
{3\over 2}C_F\right)\right.\nonumber\\
\qquad & &+\,
\left\{f_q(x_{bj},\mu)\otimes P^{\,in,(0)}_{qq}+ f_g(x_{bj},\mu)\otimes P_{qg}^{\,in,(0)}
)\right\} D_q(z_f,\mu)\,\nonumber\\
\qquad & & +\;
 f_q(x_{bj},\mu)\left\{P_{qq}^{\,out,(0)}\otimes  D_q(z_f,\mu)+
P_{gq}^{\,out,(0)}\otimes  D_g(z_f,\mu)\right\}
\Bigg] \; ,
\label{asymunpol}
\eeq
where $\alpha_{em}$ is the electromagnetic coupling constant, 
$e_q$ the fractional quark charge, and $C_F=4/3$. 
${\cal A}_1$ is defined in (\ref{A3}). In Eq.~(\ref{asymunpol}), 
$ f_q(x,\mu)$ and $ f_g(x,\mu)$ denote, respectively, the unpolarized
quark and gluon distribution functions for the proton at scale $\mu$, and
$D_q(z,\mu)$ and $ D_g(z,\mu)$ are quark and gluon fragmentation
functions for the produced hadron. Furthermore, for the unpolarized
cross section, the $P_{ij}^{\,in,(0)}$ and $P_{ij}^{\,out,(0)}$ are identical, 
and they are equal to the customary LO unpolarized splitting functions:
$P_{ij}^{\,in,(0)}=P_{ij}^{\,out,(0)}\equiv P_{ij}^{(0)}$, where
\beq
P_{qq}^{(0)}(x)&=&C_F\left[ {1+x^2\over (1-x)_+} +
{3\over 2}\delta(1-x)\right]\; ,\nonumber\\
P_{gq}^{(0)}(x)&=&C_F{1+(1-x)^2\over x} \; ,\nonumber\\
P_{qg}^{(0)}(x)&=&T_R\,\left[ x^2+(1-x)^2 \right]\; .
\label{split_unpol}
\eeq
Here, $T_R=1/2$, and the ``+''-prescription in the first line acts in 
an integral from $x$ to 1 as
\begin{equation}
\int_x^1 dy \frac{f(y)}{(1-y)_+}=
\int_x^1 dy \frac{f(y)-f(1)}{1-y}+f(1)\ln(1-x)\; ,
\end{equation} 
for any suitably regular function $f$. Finally, 
the symbol $\otimes$ denotes convolutions of the forms
\beq
&& f_q(x_{bj},\mu)\otimes P_{qq}^{\,in (0)} \equiv \int_{x_{bj}}^1{dx\over x} 
f_q(x,\mu) P_{qq}^{\,in,(0)}\left( \frac{x_{bj}}{x}\right)\; ,\nonumber \\
&&P_{qq}^{\,out,(0)}\otimes  D_q(z_f,\mu)\equiv \int_{z_f}^1{dz\over z} 
P_{qq}^{\,out,(0)}(z) D_q\left(\frac{z_f}{z},\mu\right) \; .\label{conv}
\eeq

For the other processes listed in (\ref{eq1.1}), the asymptotic parts are obtained 
in the same manner from the cross sections given in Appendix A. They can be 
cast in the form~(\ref{asymunpol}) with the following replacements:

\noindent
(ii) $e\vec{p}\to e\vec{\Lambda} X$:
\beq
f_q\to\Delta  f_q,\qquad  f_g\to\Delta  f_g,\qquad  D_q\to
\Delta  D_q,\qquad  D_g\to\Delta  D_g\; ,
\label{repl1}
\eeq
where the $\Delta f_q$, $\Delta f_g$ and $\Delta  D_q$, $\Delta  D_g$
denote the helicity-dependent quark and gluon parton densities and 
fragmentation functions, respectively (for definitions, see~\cite{trrev}). 
Furthermore,
\beq
P_{ij}^{\,in,(0)} \to \Delta P_{ij}^{(0)} \; , \quad 
P_{ij}^{\,out,(0)} \to \Delta P_{ij}^{(0)} \; ,
\eeq
with 
\beq
\Delta P_{qq}^{(0)}(x)=P_{qq}^{(0)}(x)\; , \quad 
\Delta P_{gq}^{(0)}(x)=C_F\left[ 2-x \right]\; , \quad 
\Delta P_{qg}^{(0)}(x)=T_R\left[2x-1 \right] \; . \label{split_Lpol}
\label{process2}
\eeq

\noindent
(iii) $e{p}^\uparrow\to e{\Lambda}^\uparrow X$:
\beq
& &{\cal A}_1\to \sinh^2\psi\cos(\Phi_A-\Phi_B-2\phi) \; ,\nonumber\\
& &  f_q\to\delta  f_q \; ,\qquad  f_g\to 0 \; ,\qquad  D_q\to
\delta D_q \; ,\qquad  D_g\to 0 \;,\nonumber\\
& &P_{qq}^{\,in,(0)}\to \delta P_{qq}^{(0)} \; ,
\qquad P_{qq}^{\,out,(0)}\to \delta P_{qq}^{(0)} \; .
\label{process3}
\eeq
Here, the $\delta f_q$ and $\delta D_q$ denote the transversity
distribution and fragmentation functions, respectively. As is well-known,
gluons do not contribute to transversity at leading twist~\cite{trrev}. 
The LO transversity splitting function is given by
\beq
\delta P_{qq}^{(0)}(x)=C_F\left[ {2x\over (1-x)_+} +{3\over 2}\delta(1-x)\right].
\label{split_Tpol}
\eeq

\noindent
(iv) $\vec{e}\vec{p}\to e{\Lambda} X$:
\beq
& &{\cal A}_1\to{\cal -A}_6 \; ,\qquad 
 f_q\to\Delta  f_q \; ,\qquad  f_g\to\Delta  f_g \; ,\nonumber\\
& &P_{ij}^{\,in,(0)}\to \Delta P_{ij}^{(0)} \; ,
\qquad P_{ij}^{\,out,(0)}\to P_{ij}^{(0)} \; .
\label{process4}
\eeq

\noindent
(v) $\vec{e}{p}\to e\vec{\Lambda} X$:
\beq
& &{\cal A}_1\to{\cal -A}_6 \; ,\qquad  D_q\to
\Delta  D_q \; ,\qquad  D_g\to\Delta  D_g \; ,\nonumber\\
& &P_{ij}^{\,in,(0)}\to  P_{ij}^{(0)} \; ,\qquad P_{ij}^{\,out,(0)}\to 
\Delta P_{ij}^{(0)} \; .
\label{process5}
\eeq

\subsection{NLL resummed cross section \label{sec2.4}}

Resummation of the small-$q_T$ logarithms is achieved in 
impact-parameter ($b$) space, where the large
corrections exponentiate. In $b$-space, the leading logarithms
$\alpha_S^k\,\ln^{2k-1}\left(Q^2 /q_T^2\right)/q_T^2$ turn into
$\alpha_s^k \ln^{2k}(b Q/b_0)$, where $b_0=2 {\mathrm{e}}^{-\gamma_E}$ with 
$\gamma_E$ being the Euler constant. Subleading logarithms are down 
by one or more powers of $\ln(b Q/b_0)$. In the following, we will present the
resummation formulas for the processes (i)-(v) in~(\ref{eq1.1})
we are interested in, based on the CSS formalism~\cite{CSS85}
(see also~\cite{MOS96,NSY00,yuan}). We will make some standard choices~\cite{CSS85} 
for the parameters and scales in that formalism, which render the
resulting expressions as simple and transparent as possible. 

For the unpolarized resummed cross one has
\beq
{d\sigma^{res} \over dx_{bj} dz_f dQ^2 dq_T^2 d\phi}
={\pi\alpha_{em}^2\over 2 x_{bj}^2 S_{ep}^2}\,
{\cal A}_1\int{d^2\vec{b}\over (2\pi)^2} \, {\mathrm{e}}^{i\vec{b}\cdot\vec{q}_T}\,
W(b,Q,x_{bj},z_f) \; , 
\label{resum}
\eeq
where the $b$-space expression for the resummed cross section $W$ is given by
\beq
W(b,Q,x_{bj},z_f) =\!\!\!\!\!\!
\sum_{\begin{array}{c}
j=q,\bar{q} \\
i,k=q,\bar{q},g
\end{array}}
\!\!\!\!\! \!\!\!\!e_j^2 \, \left[ f_i (x_{bj},b_0/b)\otimes C_{ji}^{\,in} \right]\, 
\times\, {\mathrm{e}}^{S(b,Q)} \times
\left[C_{kj}^{\,out}\otimes D_k(z_f,b_0/b) \right] \; ,
\label{Wbspace}
\eeq
the convolutions being defined as in Eqs.~(\ref{conv}).  As indicated, 
the scale in the parton distributions is given by $\mu=b_0/b$. We will
return to this in the next subsection.

The Sudakov form factor $S(b,Q)$ in Eq.~(\ref{Wbspace}) reads
\beq
S(b,Q)=-\int_{b_0^2/b^2}^{Q^2}{dk_T^2\over k_T^2}
\left[A\left(\alpha_s(k_T)\right)\ln\left({Q^2\over k_T^2}\right) 
+ B\left(\alpha_s(k_T)\right) \right] \; ,
\label{Sudakov}
\eeq
where the functions $A$ and $B$ have perturbative expansions of the form
\beq
A\left(\alpha_s\right)=
\sum_{k=1}^\infty A_k\left({\alpha_s\over\pi}\right)^k \; , \quad
B\left(\alpha_s\right)
=\sum_{k=1}^\infty B_k\left({\alpha_s\over\pi}\right)^k.
\eeq
For the resummation at NLL, one needs the coefficients 
$A_{1,2}$ and $B_1$, which read~\cite{DS84,KT82}: 
\beq 
A_1=C_F\; , \quad  A_2=
\frac{1}{2} \; C_F  \left[
C_A \left( \frac{67}{18} - \frac{\pi^2}{6} \right)
- \frac{5}{9} N_f \right] \; ,  \quad B_1 = -\frac{3}{2}C_F \; , 
\eeq

The functions $C_{ij}^{\,in}$ and $C_{ij}^{\,out}$ in Eq.~(\ref{Wbspace}) are also
perturbative. Their expansions read
\beq
C^{\,in/out}_{ij}(x,\alpha_s(\mu))&=&
\sum_{k=0}^\infty C^{\,in/out,(k)}_{ij}(x) 
\left({\alpha_s(\mu)\over \pi}\right)^k \; .
\label{C-func}
\eeq
Here $\mu$ is a renormalization scale of order $Q$. 
The LO coefficients are given by
\beq
C^{\,in,(0)}_{qq'}(x)&=&\delta_{qq'}\delta(1-x) \; ,\qquad
C^{\,out,(0)}_{qq'}(z)=\delta_{qq'}\delta(1-z) \; ,\nonumber\\
C^{\,in,(0)}_{qg}&=&C^{\,out,(0)}_{gq}=0 \; . 
\eeq

The first-order coefficients $C^{\,in/out,(1)}_{ij}$ could be obtained by 
expanding Eq.~(\ref{Wbspace}) to $O(\alpha_s)$, performing an inverse 
Fourier transform, and comparing to the fixed-order (${\cal O}(\alpha_s)$) 
$q_T$ distribution. The asymptotic part we have given
in Eq.~(\ref{asymunpol}) alone is not sufficient for this because it does 
not contain the contributions $\propto \delta^2(\vec{q}_T)$
at ${\cal O}(\alpha_s)$. Fortunately, the coefficients are known in the
literature, and have a very transparent structure in the 
$\overline{\mathrm{MS}}$ scheme~\cite{NSY00,Nad} (see
also work on the coefficients in the related  Drell-Yan case 
in~\cite{AEGM84,DS84,dG01}). One has:
\beq
&&C^{\,in,(1)}_{ij}(x)= -{1\over 2}P^{\,in,\epsilon}_{ij}(x)+\delta_{ij}
{\cal C}_{\delta}\delta(1-x)
\; , \label{cineps} \\
&&C^{\,out,(1)}_{ij}(z)= -{1\over 2}P^{\,out,\epsilon}_{ij}(z)
+\ln (z)P_{ij}^{out,(0)}(z)+\delta_{ij}
{\cal C}_{\delta}\delta(1-z) \; .
\label{couteps}
\eeq
Here the $P^{\,in,\epsilon}_{ij}(x)$, $P^{\,out,\epsilon}_{ij}(x)$
are the terms $\propto \epsilon$ in the LO splitting functions
in $d=4-2\epsilon$ dimensions, $P_{ij}^{\, in/out, (0)}(x,\epsilon)$, 
and are defined through
\beq
P_{ij}^{\, in/out, (0)}(x,\epsilon)=P_{ij}^{in/out,(0)}(x)+\epsilon
P_{ij}^{\,in/out,\epsilon}(x) \; .
\eeq
The usual splitting functions in $d=4$ dimensions have been 
given above in Eqs.~(\ref{split_unpol}). One has~\cite{ERT}
\beq
\hspace*{-3mm}
P_{qq}^{\,\epsilon}(x)=-C_F\left[ 1-x-\frac{1}{2}\delta(1-x)\right]
\; , \;\;\; P_{qg}^{\,\epsilon} (x)=-2T_Rx(1-x) \; ,
\;\;\; P_{gq}^{\,\epsilon} (x)=-C_Fx \; .
\label{epsilonunpol}
\eeq
Furthermore, the coefficient ${\cal C}_{\delta}$ in Eqs.~(\ref{cineps}),(\ref{couteps})
arises from hard virtual corrections. It is therefore only present for the case 
$ij=qq$, where ${\cal C}_{\delta}=-7C_F/4$. Finally, for the ``final-state''
coefficient $C^{\,out,(1)}_{ij}$ there is an extra contribution
$\propto \ln z$~\cite{NSY00,Nad} times the LO splitting function. 
It is of kinematic origin and contributed by the phase space for collinear gluon 
radiation off the quark leg in the final state~\cite{SV}.  
 
With this, the formulas for the resummation of the unpolarized
SIDIS cross section in $b$-space are complete. It is now straightforward
to extend them to the various spin-dependent cross sections in~(\ref{eq1.1}). 
The Sudakov form factors are entirely related to soft-gluon emission and 
hence are spin-independent and the same in each case~\cite{yuan}. 
For each process, we first need to make the appropriate replacements 
as described above in Eqs.~(\ref{repl1})-(\ref{process5}).  
In addition, we obtain the respective coefficients $C_{ij}^{\,in/out,(1)}$ from 
Eqs.~(\ref{cineps}),(\ref{couteps}), substituting appropriately all
splitting functions, and also the  
$P_{ij}^{\,in/out,\epsilon}$ by the $\propto\epsilon$ terms in the 
$d=4-2\epsilon$-dimensional versions of the LO splitting functions
that~(\ref{repl1})-(\ref{process5}) 
direct us to use. For this we need the $\propto\epsilon$
terms in the spin-dependent splitting functions. For longitudinal polarization we 
have~\cite{Vogelsang95}
\beq
&&\Delta P_{qq}^{\,\epsilon}(x)=-C_F\left[ 1-x-\frac{1}{2}\delta(1-x)
\right] \; , \nonumber \\ 
&&\Delta P_{qg}^{\,\epsilon} (x)=-2T_R(1-x) \; ,
\;\; \Delta P_{gq}^{\,\epsilon} (x)=2C_F(1-x) \; .
\label{epsilonpol}
\eeq
In case of transversity, the splitting function in $4-2\epsilon$ dimensions
is for $x<1$ identical to that in four dimensions~\cite{Vogelsang98}, so that
\beq
\delta P_{qq}^{\,\epsilon} (x) = \frac{C_F}{2}\delta(1-x) \; .
\label{epsilonT}
\eeq
The coefficient ${\cal C}_{\delta}$ for the case $ij=qq$ 
in Eqs.~(\ref{cineps}),(\ref{couteps}) is the same for all polarized cases, 
since it comes from virtual diagrams. 

Knowledge of the coefficients $A_{1,2}$, $B_1$ and $C_{ij}^{(0),(1)}$ in 
Eqs.~(\ref{resum}),(\ref{Wbspace})
is sufficient for the resummation to NLL accuracy. For consistency,
we also need to use the two-loop expression for the strong running coupling, 
and the NLO evolution for the parton densities and fragmentation functions.
Let us finally give an explicit formula for the NLL expansion of
the Sudakov form factor. Defining 
\beq
L=\ln\left(1+Q^2b^2/b_0^2 \right) \; ,
\label{logdef}
\eeq
and choosing the renormalization scale $\mu$, we have~\cite{FNR99}:
\beq
S(Q,b)={1\over \alpha_s(\mu)}f_0(\alpha_s(\mu)L)+f_1(\alpha_s(\mu)L) \; ,
\label{PSudakov}
\eeq
where
\beq
f_0(y)&=&{A_1\over \pi\beta_0^2}\left[ \beta_0 y+\ln(1-\beta_0y)\right],\\
f_1(y)&=&{A_1\beta_1\over \pi\beta_0^3}\left[ {1\over 2}\ln^2(1-\beta_0y)+
{\beta_0y\over 1-\beta_0 y}
+{\ln(1-\beta_0y)\over 1-\beta_0y}\right]+ 
{B_1\over \pi \beta_0}\ln(1-\beta_0y)\nonumber\\
& &+\frac{1}{\pi \beta_0}\left[A_1\ln\left(\frac{Q^2}{\mu^2}\right)
-{A_2\over \pi\beta_0}\right]
\left[ \ln(1-\beta_0y) +{\beta_0y\over 1-\beta_0y}\right] \; ,
\eeq
with the coefficients of the QCD $\beta$-function $\beta_0=(33-2N_f)/(12\pi)$ and 
$\beta_1=(153-19 N_F)/(24 \pi^2)$. 
We observe that as $y\to 0$, $f_0(y)={\cal O}(y^2)$ and $f_1(y)={\cal O}(y)$.  
Note that in Eq.~(\ref{logdef}) 
we have followed Ref.~\cite{BCDG03} to choose $\ln\left(1+Q^2b^2/
b_0^2 \right)$ rather than $\ln\left(Q^2b^2/b_0^2 \right)$ as the large
logarithm in $b$-space. 
As far as the large-$b$ behavior and NLL resummation 
are concerned, the two are equivalent, of course. With the choice 
in~(\ref{logdef}), however, the Sudakov exponent vanishes at $b=0$, 
as it should~\cite{PP79,AEGM84,CSS85}, whereas for the other choice it gives large 
logarithms not only at large $b$ (small $q_T$) but also artificially 
at small $b$ (large $q_T$). For further discussion, see~\cite{BCDG03}.

\subsection{Evolution of parton distributions and fragmentation \\ functions
\label{sec2.5}}

As we have seen in Eq.~(\ref{Wbspace}), in the CSS formalism the parton 
distributions and fragmentation functions are evaluated at
the factorization scale $\mu=b_0/b$. As discussed in~\cite{KSV02}, 
it then becomes a great convenience to treat them in 
Mellin-moment space, because this enables one to explicitly express their 
evolution between a large scale $\sim Q$ and $b_0/b$. In this way, one avoids 
the problem normally faced in $q_T$ resummation that one needs to call 
the parton densities or fragmentation functions at scales far below their 
range of validity (see below), 
so that some sort of ``freezing'' (or related prescription) 
for handling them is required. We also anticipate that below we will choose
the impact parameter $b$ to be complex-valued, so that it becomes
desirable to separate the complex scale $b_0/b$ from the scale at which the
parton densities are explicitly evaluated. 
 
To be more specific, we write the resummed cross section 
as $W(b,Q,x_{bj},z_f)$ in Eq.~(\ref{Wbspace}) as an inverse
Mellin transform of moments in $x_{bj}$ and $z_f$: 
\beq
\hspace*{-1.5cm}W(b,Q,x_{bj},z_f) &=&\!\!\!\!\!\!
\sum_{\begin{array}{c}
j=q,\bar{q} \\
i,k=q,\bar{q},g
\end{array}}
\!\!\!\!\! \!\!\!\!e_j^2 \, \left({1\over 2\pi i}\right)^2
\int_{{{\cal C}}_N}\,dN\,\int_{{{\cal C}}_M}\,dM\, z_f^{-M} x_{bj}^{-N}\nonumber\\
&&\hspace*{1.5cm}\times \, f_i^N (b_0/b)C_{ji}^{\,in,N}  
{\mathrm{e}}^{S(b,Q)} 
C_{kj}^{\,out,M} D_k^M(b_0/b) \; ,
\label{InverseMellin}
\eeq
where the moment of each function is defined as
\beq
f^N_i(\mu)\equiv \int_0^1 \,x^{N-1}f_i(x,\mu) \; ,\qquad
C^{\,in,N}_{ji}(\alpha_s(\mu))\equiv 
\int_0^1\,x^{N-1}C^{\,in}_{ji}(x,\alpha_s(\mu)) \; ,
\eeq
and so forth. The integrations over $N$ and $M$ in Eq.~(\ref{InverseMellin})
are over contours in the complex plane, to be chosen in such a way that they lie
to the right of the rightmost singularities of the parton distributions and the 
fragmentation functions. In moment space, it is possible to express
the moments $f_i^N(b_0/b)$ and $D_k^M(b_0/b)$ by the corresponding moments at 
scale $\sim Q$. The evolution ``factor''  between the scales $b_0/b$ and $Q$
(which in general is a matrix because of singlet mixing) can be written 
in closed analytical form and can be expanded to NLL accuracy in $L$. One
then only needs the parton densities and fragmentation functions at scale $Q$.
For further details, see~\cite{KSV02}. 

\subsection{Inverse Fourier transform and non-perturbative corrections \label{sec32}}

As shown in Eq.~(\ref{resum}), the resummed cross section in $q_T$-space
is obtained by an inverse Fourier transform of $W(b,Q,x_{bj},z_f)$.
This involves an integration over $b=|\vec{b}|$ from 
0 to $\infty$. Because of the Landau pole of the perturbative strong
coupling in the Sudakov exponent~(\ref{Sudakov}) at $k_T=\Lambda_{\mathrm{QCD}}$,
this integration is ill-defined. In the expansion~(\ref{PSudakov}) 
for the Sudakov form factor, the singularity occurs at $\beta_0\alpha_s(\mu)L=1$.
In other words, the $b$-integration extends over both perturbative 
($b\ll 1/\Lambda_{QCD}$) and non-perturbative ($b\leq 1/\Lambda_{QCD}$ or even 
larger) regions. In order to define a perturbative resummed cross section, a
prescription for the $b$ integration is required that avoids the Landau 
pole. The ambiguity in the perturbative series implied by the presence
of the Landau pole corresponds to a non-perturbative correction. One may
therefore use resummation to examine the structure of non-perturbative
corrections in hadronic cross sections~\cite{SWV}. Typically, for the $q_T$-differential
Drell-Yan or SIDIS cross sections, one expects Gaussian non-perturbative
effects in $b$-space~\cite{CSS85,Tafat}: 
\beq
{\mathrm{e}}^{S(b,Q)}\to {\mathrm{e}}^{S(b,Q)-gb^2} \; ,
\label{Gaussian}
\eeq
where $g$ is a coefficient that may be determined by comparison to data. 
It will depend in general on $Q$ and on $x_{bj}$ and $z_f$. 
This non-perturbative Gaussian form factor may be thought of to partly 
incorporate (or, at least, mimic) the effect of smearing due to the partons' 
intrinsic transverse momenta. 

Several prescriptions for treating the inverse Fourier transform 
have been proposed in the literature. The earliest one, known as
``$b^{\ast}$-prescription''~\cite{CSS85,DS84}, is to prevent $b$ 
from becoming larger than a certain $b_{\rm max}$ in the Sudakov form
factor by replacing $b\to b^*=b/\sqrt{1+b^2/b_{\rm max}^2}$. 
This evidently introduces a new parameter, $b_{\rm max}$. The prescription
was refined recently in~\cite{Kony}. Extensive phenomenological 
studies using this prescription and a non-perturbative term as 
in~(\ref{Gaussian}) have been carried out in particular 
for the $q_T$-distribution in the Drell-Yan process~\cite{DS84,Kony,DYbstar}.
The method has also been used in studies of the $q_T$-distribution in 
unpolarized SIDIS~\cite{NSY00,Nad,Nad1}. 

Another method, proposed in~\cite{LSV} and applied to the Drell-Yan
cross section in~\cite{KSV02}, is to deform the $b$ integration
in Eq.~(\ref{resum}) into a contour in 
the complex $b$-plane. In this way, the Landau pole
is avoided since it lies far out on the real-$b$ axis. This procedure 
does not introduce any new parameter and is identical to the original 
$b$-integration in~(\ref{resum}) for any finite-order expansion of
the Sudakov exponent. For all details of the complex-$b$ prescription,
see~\cite{KSV02}. This method was also recently
used in a study of the transverse-momentum distribution of 
Higgs bosons at the LHC~\cite{BCDG03,KSV04}. In the present paper we will use both 
the $b^{\ast}$ and 
the complex-$b$ methods and compare the results and their dependence
on the non-perturbative parameters chosen. We note that for the 
complex-$b$ method the $b$-integral converges even for $g=0$ in
Eq.~(\ref{Gaussian}), while for the $b^{\ast}$ prescription a 
(non-perturbative) suppression of the integrand is required~\cite{KSV02}. 
We finally mention that other methods for dealing with the behavior of 
the resummed exponent at large $b$ have been discussed and 
used~\cite{Qiu}. Also resummations directly in $q_T$-space have 
been studied~\cite{ellis}. 

\subsection{Matching to finite order \label{match}}

In order to obtain an adequate theoretical description also
at large $q_T\sim Q$, we ``match'' the resummed cross section 
to the fixed-order (LO, ${\cal O}(\alpha_s)$) one. 
This is achieved by subtracting from the resummed expression 
in Eq.~(\ref{resum}) its expansion to ${\cal O}(\alpha_s)$,
\beq
{\pi\alpha_{em}^2\over 2 x_{bj}^2 S_{ep}^2}\,
{\cal A}_1\int{d^2\vec{b}\over (2\pi)^2} \, {\mathrm{e}}^{i\vec{b}\cdot
\vec{q}_T}\,\left[ 
W(b,Q,x_{bj},z_f) -W(b,Q,x_{bj},z_f)|_{{\cal O}(\alpha_s)} \right] \; , 
\label{resummatch}
\eeq
and then adding the full ${\cal O}(\alpha_s)$ cross section, given 
by Eq.~(\ref{asymptotic}). We use this matching procedure, which avoids any 
double-counting of higher perturbative orders, in an identical way for the 
various polarized cross sections.

\section{Numerical analysis for $\vec{e}\vec{p}\to e\pi X$ \label{sec3}}

In order to study the effect of resummation on the polarized SIDIS cross 
section and spin asymmetry, we will perform a numerical calculation for 
one case considered in~(\ref{eq1.1}), $\vec{l}\vec{p}\to l\pi X$.
This calculation is relevant for the ongoing COMPASS experiment and for 
experiments at the proposed polarized $ep$ collider eRHIC.
As typical values of the kinematical parameters, we choose 
$S_{ep}=10^4$ GeV$^2$, $Q^2=100$ GeV$^2$, $x_{bj}=0.012$ for eRHIC
and $S_{ep}=300$ GeV$^2$, $Q^2=10$ GeV$^2$, $x_{bj}=0.04$ for COMPASS.
Clearly, these choices are only meant to represent the typical
kinematics; future detailed comparisons to experimental data will
likely require to study a much broader range of values.
We integrate over $z_f>0.2$. Our two sets of the parameters give
an identical $\,\cosh\psi\,$ in Eqs.~(\ref{eq2.cosh}) and (\ref{A3}).  
For the NLO parton densities and the fragmentation functions,
we use the GRV unpolarized distributions~\cite{GRV98}, the GRSV ``standard'' 
polarized distributions~\cite{GRSV00}, and the pion fragmentation functions by
Kretzer~\cite{Kretzer00}. We have each of these available in Mellin-moment
space, which is very helpful in the light of the discussion in Subsec.~\ref{sec2.5}.

In addition, we will use a simple Gaussian non-perturbative factor as
in Eq.~(\ref{Gaussian}). We emphasize again the more illustrative
character of our study and will therefore choose only some representative
values for $g$, in order to investigate the sensitivity of the results to 
$g$. For the complex-$b$ prescription, we will use $g=0.6, \,0.8$
GeV$^2$
for both the eRHIC and the COMPASS cases.  
The value $g=0.8$~GeV$^2$ was
found in~\cite{KSV02} in the context of the ``joint'' resummation
for $Z$ production at the Tevatron. This value therefore applies to the scale 
$M_Z$. $g$ is expected to have logarithmic dependence on
$Q$~\cite{DYbstar} and should be smaller at lower energy. This motivates
our lower choice of $g=0.6$~GeV$^2$. For the $b^{\ast}$ 
method, we will choose $b_{\rm max}=1/(\sqrt{2}$~GeV$)$ and the values 
$g=0.8, \,1.3$~GeV$^2$ for the eRHIC case and
$g=0.4, \,0.8$~GeV$^2$ for COMPASS.  
We note that in the analysis~\cite{NSY00,Nad} of the HERA 
data~\cite{h1,zeus} for SIDIS observables, a detailed phenomenological 
study of non-perturbative effects was performed, using the $b^{\ast}$ 
prescription. From the energy 
flow observable in SIDIS it was found that the non-perturbative effects
appear to become larger at small $x_{bj}$. If we boldly assumed that 
the non-perturbative form of~\cite{NSY00,Nad} also applied 
for the kinematics and the observable we are considering here, we would 
find $g\approx 0.4$~GeV$^2$ for the COMPASS situation, and $g\approx 
1.3$~GeV$^2$ at eRHIC.  We are also interested in the comparison
of the two methods with the same non-perturbative Gaussian, hence the
additional choice of $g=0.8$~GeV$^2$.
Finally, we shall always use the same values of $g$ 
in the unpolarized and the polarized cases. This assumption
is motivated by the fact that the perturbative Sudakov form factors
are independent of polarization. 
\begin{figure}[t!]
\begin{center}
\vspace*{0.8cm}
\hspace*{-5mm}
\epsfig{figure=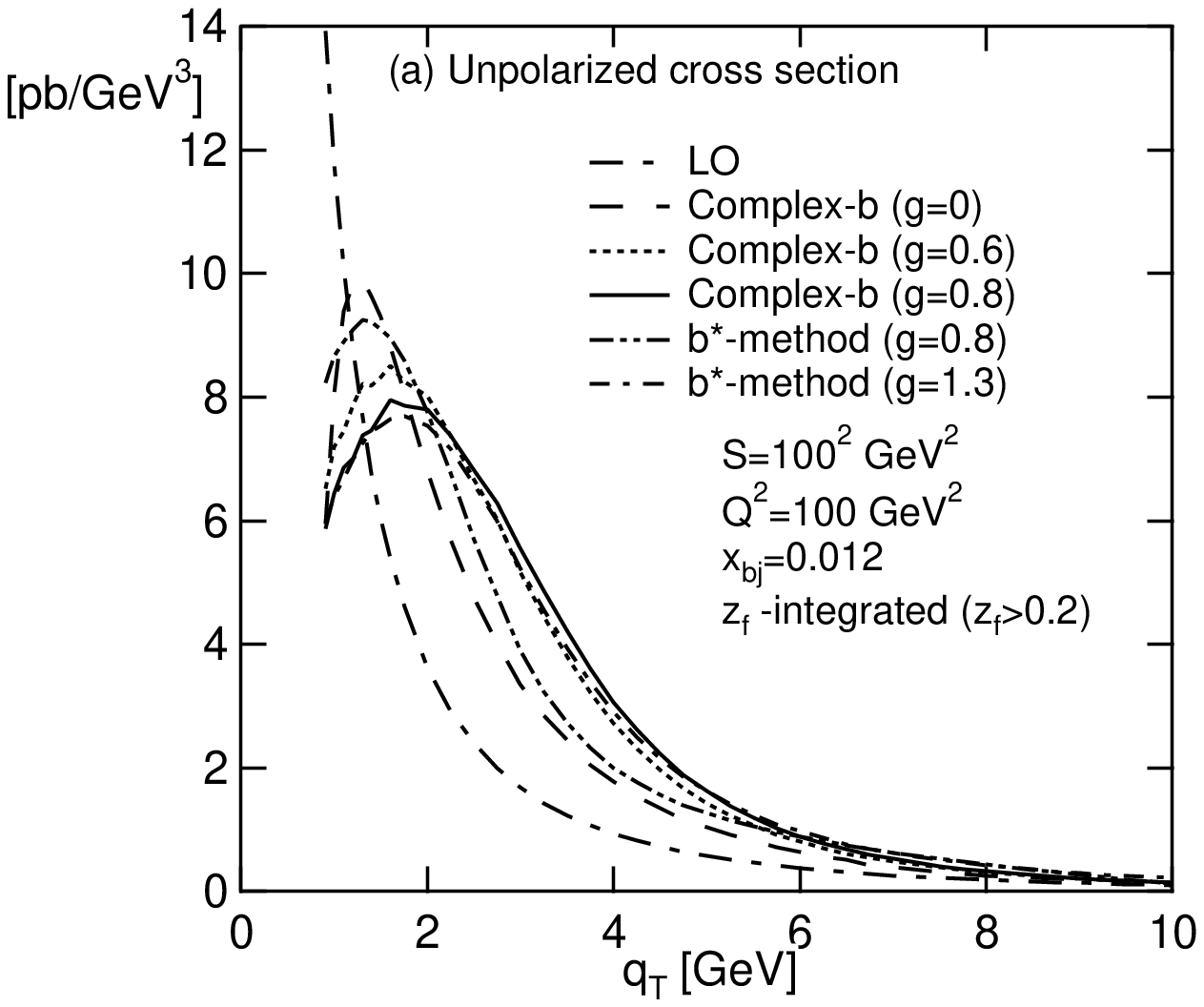,width=0.47\textwidth}
\hspace*{5mm}
\epsfig{figure=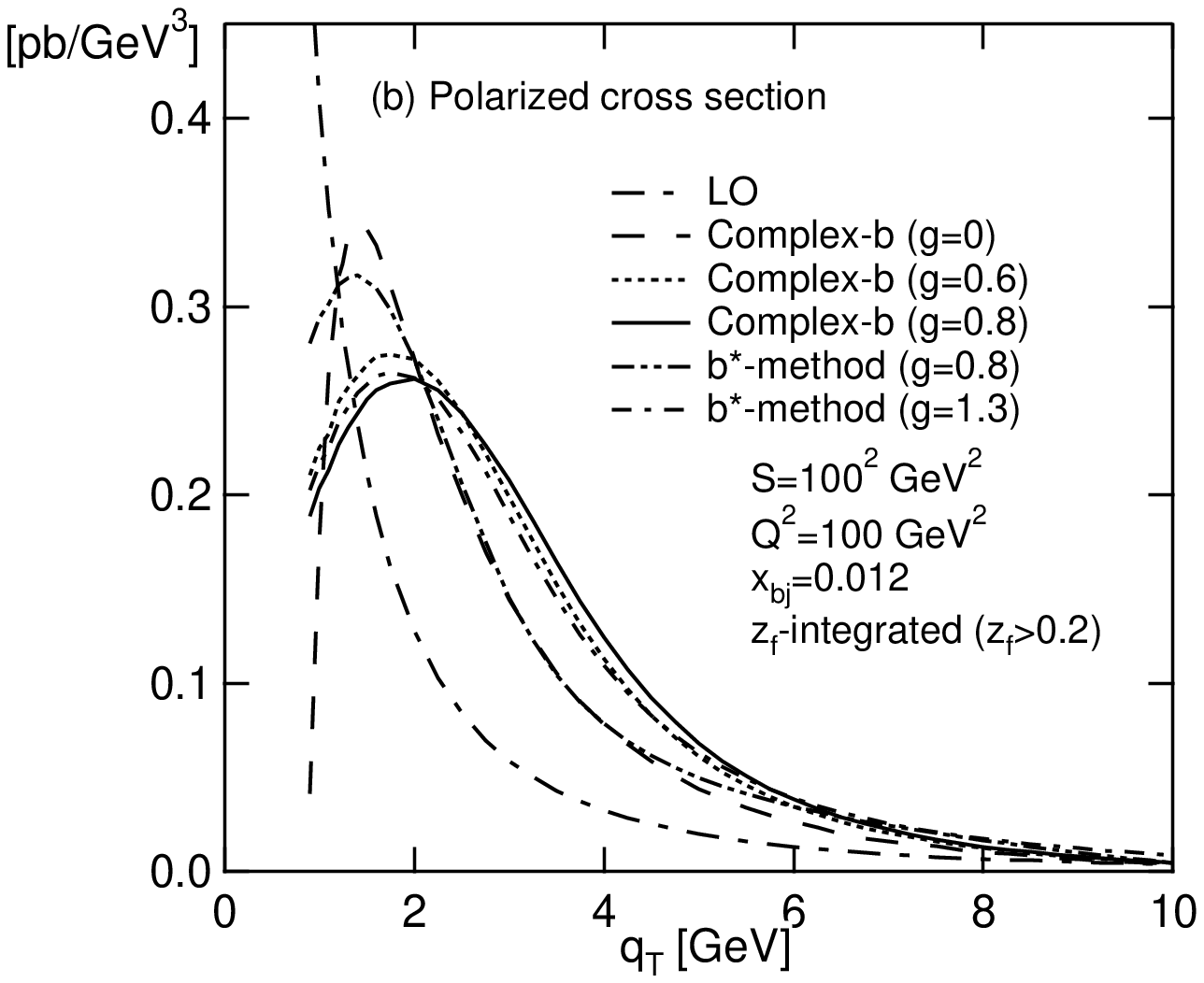,width=0.47\textwidth}
\epsfig{figure=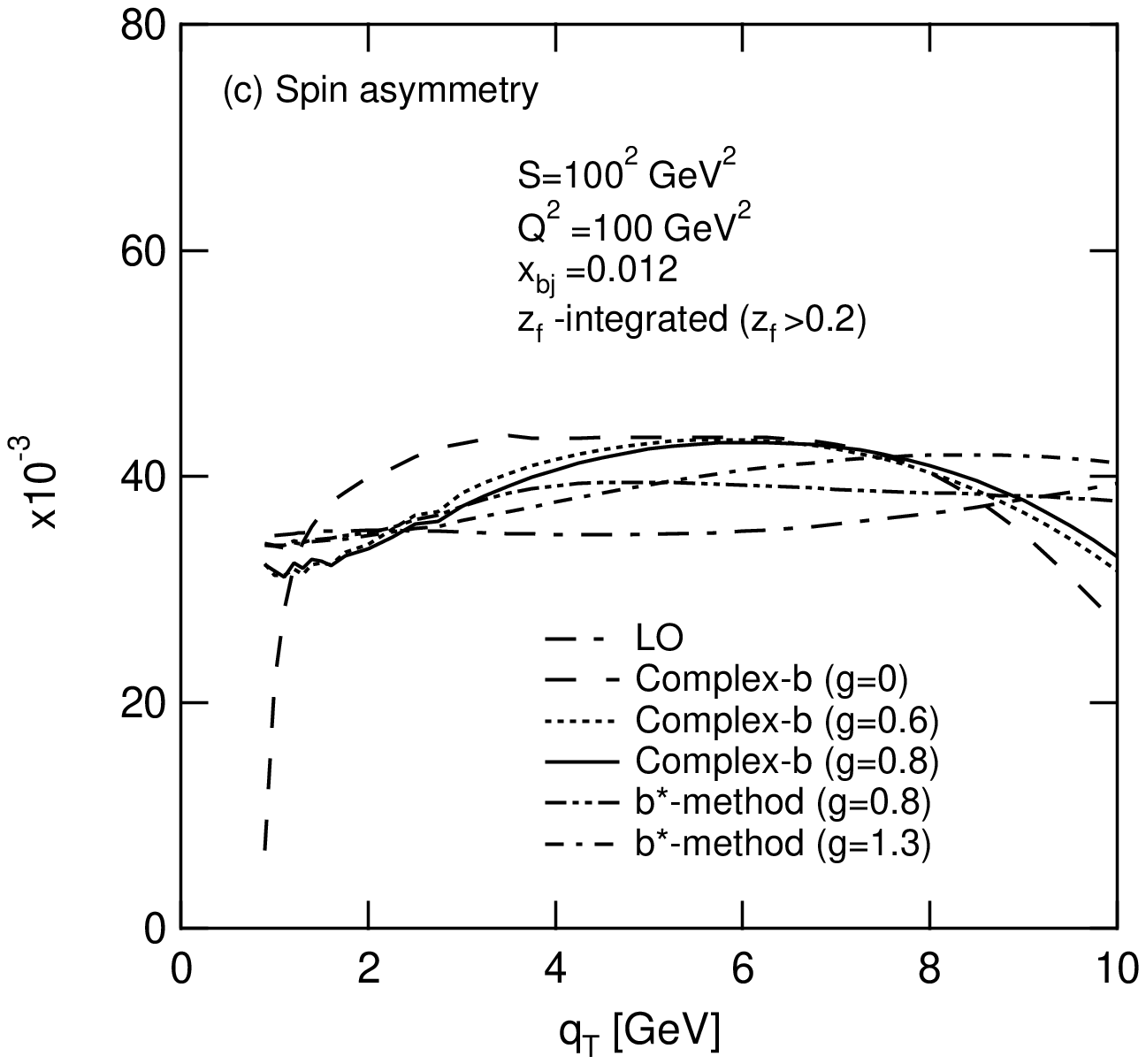,width=0.45\textwidth}
\end{center}
\vspace*{-.5cm}
\caption{(a) Unpolarized SIDIS cross section for eRHIC kinematics.
We show the fixed-order (LO) result, and resummed results for the 
complex-$b$ method with non-perturbative parameters $g=0$ and
$g=0.6,\,0.8$~GeV$^2$, and for the $b^{\ast}$ 
method with $b_{\rm max}=1/(\sqrt{2}$~GeV$)$ and $g=0.8, \, 1.3$~GeV$^2$.
(b) Same for the longitudinally polarized case. (c) Spin asymmetries
corresponding to the various cross sections shown in~(a) and~(b).
\label{fig1}}
\vspace*{0.cm}
\end{figure}

Figures~\ref{fig1}~(a) and~(b), respectively, show the 
unpolarized and polarized cross sections
\beq
{1\over 2\pi}\int_{0.2}^{z_f^{max}}dz_fd\phi{d(\Delta)\sigma 
\over dx_{bj} dz_f dQ^2 dq_T d\phi}
\eeq
for eRHIC kinematics, where $z_f^{\rm max}$ is given in Eq.~(\ref{range_zf}).  
Note that we have multiplied by a factor $q_T$, as compared to the 
cross section we considered in~(\ref{asymptotic}).  
The LO cross sections rapidly diverge as $q_T\to 0$. For the matched cross
section using the complex-$b$ method with $g=0$, one obtains an enhancement 
at intermediate $q_T$ and the expected reduction at small $q_T$. The inclusion 
of the non-perturbative Gaussian form factor makes this tendency stronger.  
However, the results for different choices of the Gaussian, $g=0.6$ and
$0.8$~GeV$^2$, are not very different and just have a slightly smaller
normalization and are shifted to the right. 
\begin{figure}[t!]
\begin{center}
\vspace*{0.8cm}
\hspace*{-5mm}
\epsfig{figure=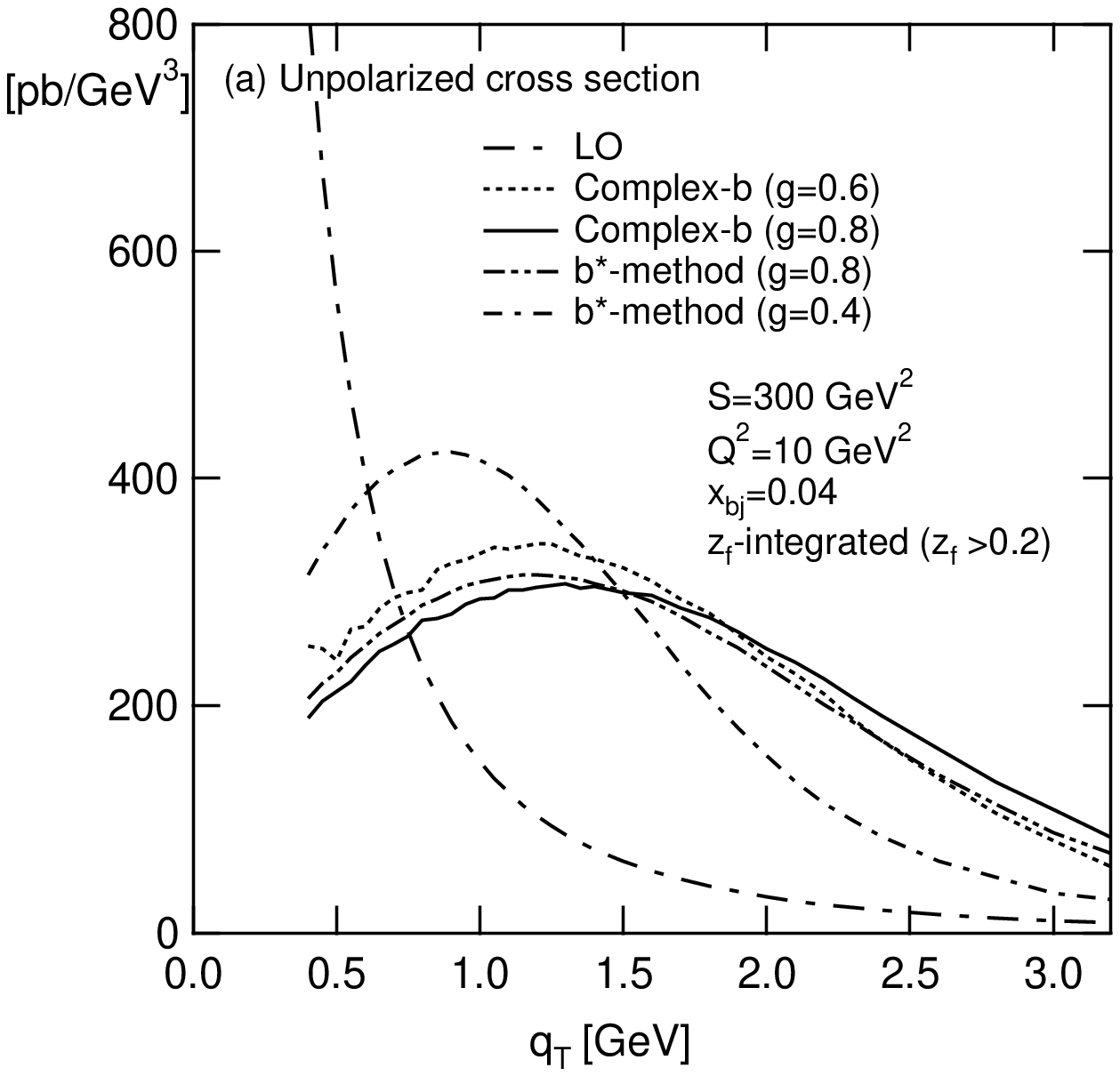,width=0.45\textwidth}
\hspace*{5mm}
\epsfig{figure=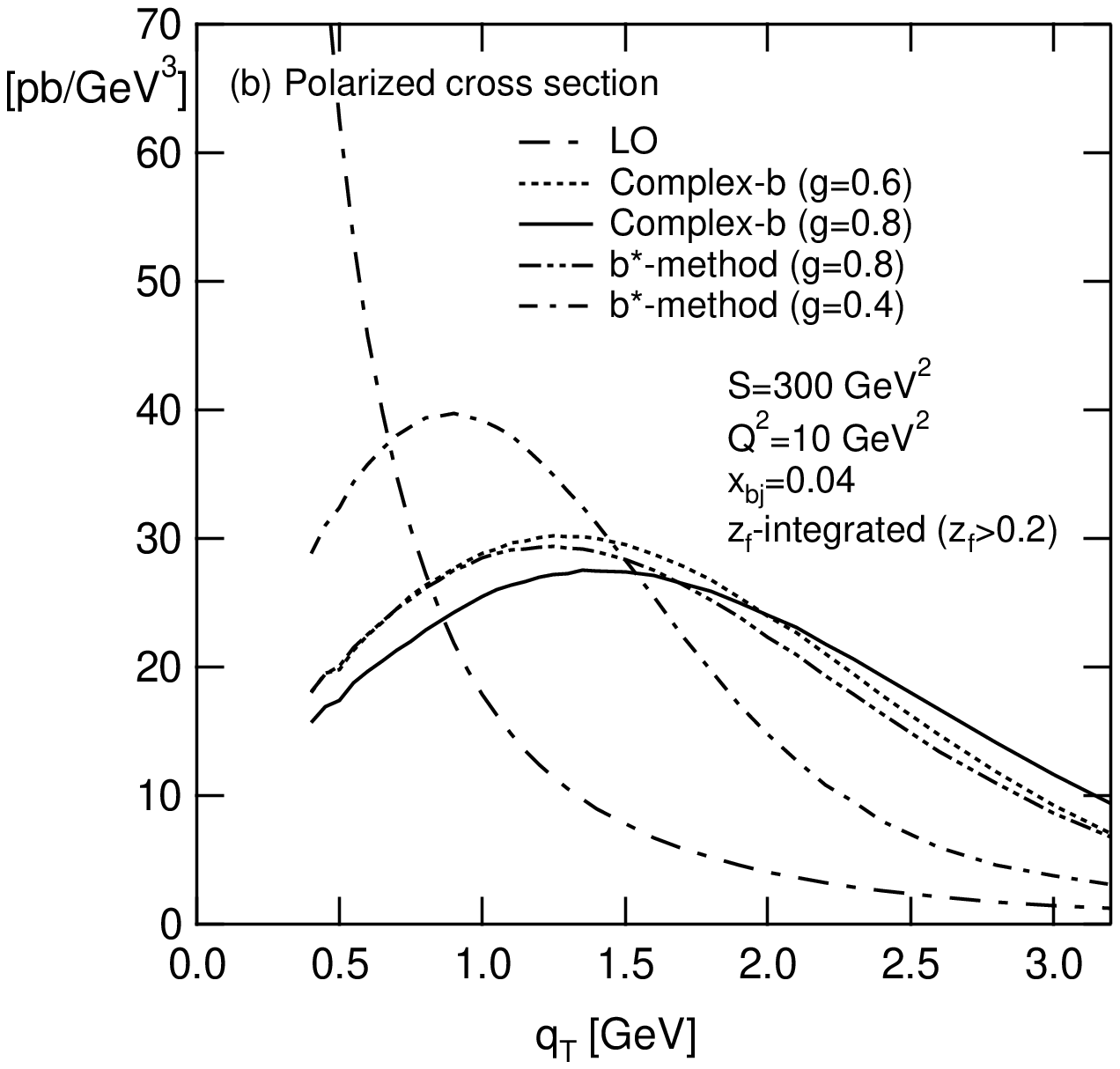,width=0.45\textwidth}
\epsfig{figure=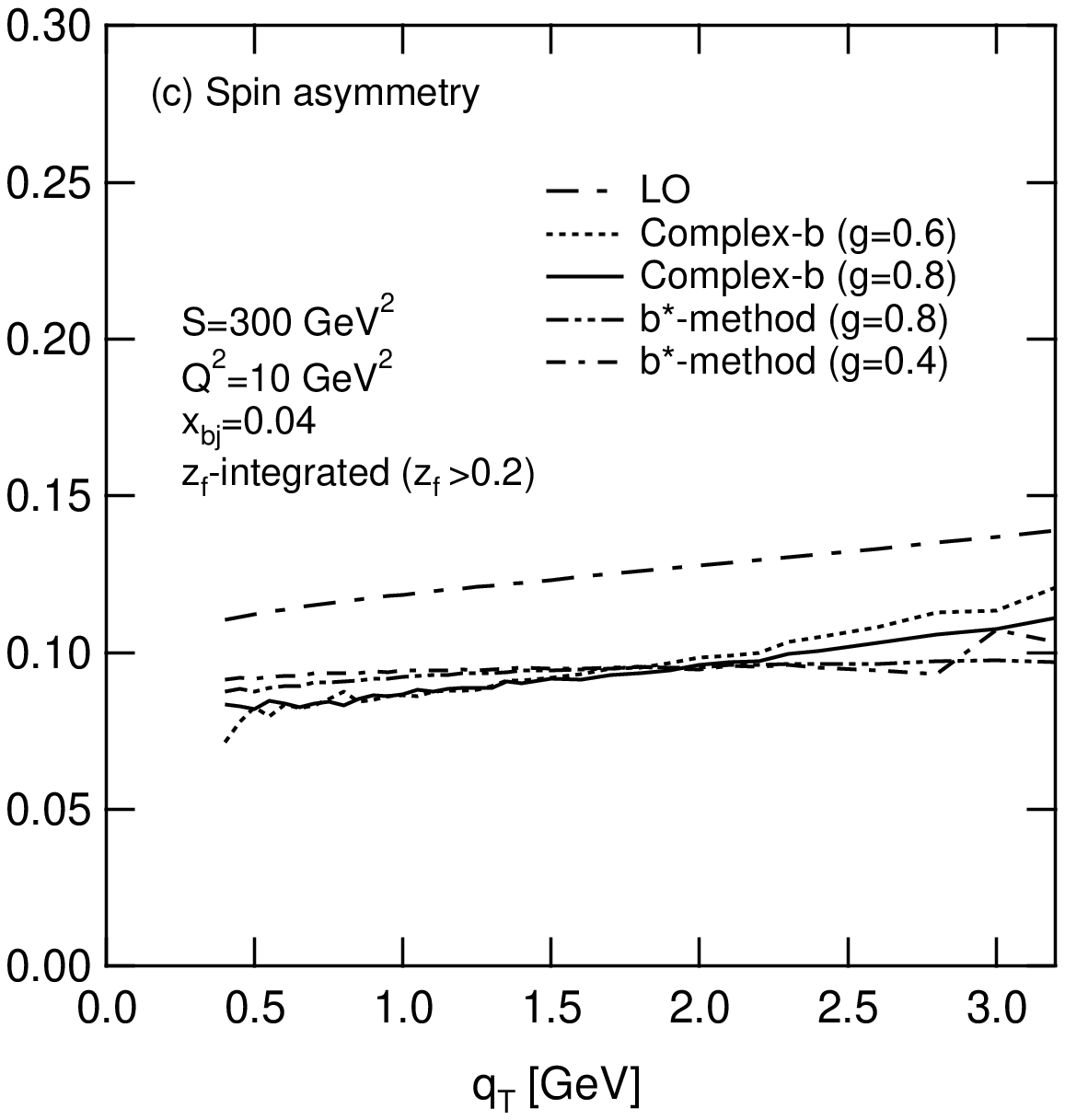,width=0.45\textwidth}
\end{center}
\vspace*{-.5cm}
\caption{Same as in Figs.~\ref{fig1}~(a)-(c), but for COMPASS
kinematics. For the $b^{\ast}$ prescription, we have chosen here 
the non-perturbative parameters 
$g=0.4, \, 0.8$~GeV$^2$. \label{fig2}}
\vspace*{0.cm}
\end{figure}
Also shown in these figures are the curves for the $b^{\ast}$ prescription 
with $b_{\rm max}=1/(\sqrt{2}$~GeV$)$ and 
$g=0.8$ and~$1.3$ GeV$^2$.  For the eRHIC case, the choice of $g=0.8$ gives a result relatively 
close to the one with $g=0$ in the complex-$b$ method,
while $g=1.3$ gives a result very close to
those with $g=0.6$ and $0.8$ in the complex-$b$ method. 
One should note that 
these resummed curves all give a very similar $q_T$-integrated cross 
section, close to the full NLO one, due to our choice of $L$ in 
Eq.~(\ref{logdef}) and our matching procedure described in the 
previous subsection~\cite{BCDG03}. 

Figure~\ref{fig1}~(c) shows the corresponding spin asymmetries, defined by the 
ratios of the polarized and unpolarized cross sections for all the curves 
shown in Figs.~\ref{fig1}~(a) and~(b).
Although the effects of resummation and the non-perturbative Gaussians
are significant in each cross section, they cancel to a large degree in the
spin asymmetry. If one looks in more detail, resummation somewhat enhances 
the asymmetry compared to LO in the range of small to intermediate $q_T$, 
where also the cross sections shown in Figs.~\ref{fig1}~(a) and~(b)
receive enhancements due to the resummation.  

Figures~\ref{fig2}~(a)-(c) show the same quantities as Figs.\ref{fig1}~(a)-(c), now
for the COMPASS kinematics described above. In this case, the complex-$b$ 
method without Gaussian smearing ($g=0$) turned out to be difficult to
control numerically at small $q_T$, and we do not show the result
for it. We find that resummation leads to a significant enhancement 
of the cross section at $q_T\geq 1$~GeV. 
In both unpolarized and polarized cross sections,
the resummed results we show, for the complex-$b$ method with $g=0.6$ and
$0.8$~GeV$^2$, and for the $b^{\ast}$ prescription with
$b_{\rm max}=1/(\sqrt{2}$~GeV$)$ and $g=0.8$~GeV$^2$,
turn out to be very similar, while the $b^{\ast}$ prescription
with $g=0.4$~GeV$^2$ gives a higher peak that is shifted to the left
compared to the other three resummed results. All these resummed results 
give a very similar spin asymmetry for the process $\vec{l}\vec{p}\to l\pi X$
for COMPASS kinematics; we find that resummation just leads to a moderate decrease 
of the asymmetry.

\section{Summary and Conclusions \label{sec4}}
We have carried out a study of the soft-gluon resummation for the 
transverse-momentum ($q_T$) distribution in semi-inclusive 
deeply-inelastic scattering. Resummation is crucial at small transverse
momenta, $q_T\ll Q$, where it takes into account large double-logarithmic
corrections to all orders in the strong coupling constant. 
We have considered all relevant 
leading-twist double-spin cross sections, focusing on the terms 
that are independent of the angle between the lepton and the hadron
planes, and have presented the resummation formulas for each. 

We have performed phenomenological studies for the process
$\vec{l}\vec{p}\to l\pi X$ at COMPASS and at a possible 
future polarized $ep$ collider, eRHIC. Here we have chosen two different
prescriptions for treating the region of very large impact parameters
in the Sudakov form factor, which is related to the onset of 
non-perturbative phenomena. We have used simple estimates for the 
non-perturbative term suggested by the resummed formula. 
Our results indicate that resummation effects as well as 
non-perturbative effects cancel to a large extent in the spin asymmetry.

\section*{Acknowledgments}
We are grateful to D.\ Boer, D.\ de Florian and F.\ Yuan for
very useful discussions and comments. 
W.V.\ thanks RIKEN, Brookhaven National Laboratory (BNL)
and the U.S.\ Department of Energy (contract number DE-AC02-98CH10886) for
providing the facilities essential for the completion of his work.
The major part of this work was done while Y.K.\ stayed at 
BNL during 2003-2004 by the support from Monbu-Kagaku-sho.
Y.K.\ is grateful to Monbu-Kagaku-sho for the financial support and 
to the BNL Physics Department, in particular to Larry McLerran, for the 
warm hospitality during his stay.


\appendix

\section{Analytic formulas for the LO cross sections}
\setcounter{equation}{0}
\renewcommand{\theequation}{A.\arabic{equation}}

Here we summarize the LO cross section formulas for the 
processes in~(\ref{eq1.1}), which were derived in
\cite{KN03}. The differential cross sections are given by
\beq
{d^5\sigma\over dx_{bj} dQ^2 dz_f dq_T^2 d\phi}
&=& {\alpha_{em}^2 \alpha_s \over 8\pi x_{bj}^2 S_{ep}^2 Q^2}
\sum_k {\cal A}_k \int_{x_{min}}^1\,{dx\over x}\int_{z_{min}}^1\,{dz\over z}\,
\left[ f\circ D\circ\widehat{\sigma}_k\right] \nonumber\\
& &\qquad\qquad\qquad\times\,\delta\left( {q_T^2\over Q^2} -
\left( {1\over \xhat} -1\right)\left({1\over \zhat}-1\right)\right) \; ,
\label{eq3.1}
\eeq
where the ${\cal A}_k$ are defined as\,\cite{KN03,MOS92}
\beq
{\cal A}_1 &=& 1+\cosh^2\psi\; ,\nonumber\\
{\cal A}_2 &=& -2\; ,\nonumber\\
{\cal A}_3 &=& -\cos\phi\sinh 2\psi\; ,\nonumber\\
{\cal A}_4 &=& \cos 2\phi\sinh^2\psi\; ,\nonumber\\
{\cal A}_6 &=& -2\cosh\psi\; ,\nonumber\\
{\cal A}_7 &=&2\cos\phi\sinh\psi\; ,\nonumber\\
{\cal A}_8 &=& -\sin\phi\sinh 2\psi\; ,\nonumber\\
{\cal A}_9 &=& \sin 2\phi\sinh^2\psi\; ,
\label{A3}
\eeq
and the summation over $k$ is $k=1,\ldots,4$ for processes (i) and
(ii) in (\ref{eq1.1}), $k=1,\ldots,4,8,9$ for process (iii), and $k=6, 7$
for processes (iv) and (v) (see below).
$\alpha_{em}=e^2/4\pi$ is the QED coupling constant, and 
we have introduced the variables
\beq
\xhat={x_{bj}\over x}\; ,\qquad \zhat={z_f\over z}\; ,
\eeq
and 
\beq
x_{min}=x_{bj}\left( 1 + {z_f\over 1-z_f}{q_T^2\over Q^2}\right)\; ,\qquad
z_{min}=z_{f}\left( 1 + {x_{bj}\over 1-x_{bj}}{q_T^2\over Q^2}\right) \; .
\label{eq3.2}
\eeq
For a given $S_{ep}$, $Q^2$ and $q_T$,
the kinematic constraints for $x_{bj}$ and $z_f$ are
\beq
& &{Q^2\over S_{ep}}<x_{bj}<1 \; ,
\label{range_zb}\\
& &0< z_f < {1-x_{bj}\over 
1-x_{bj}+x_{bj}q_T^2/Q^2} \; .
\label{range_zf}
\eeq  
Consequently, $q_T$ is limited by 
\beq
0< q_T < Q\sqrt{\left({1\over x_{bj}}-1\right)\left({1\over z_{f}}-1\right)} \; .
\eeq
In the hadron frame, the transverse momentum, $p_T$, of $p_B^\mu$ obeys
\beq
p_{T}=z_f q_T <
z_fQ\sqrt{\left({1\over x_{bj}}-1\right)\left({1\over z_{f}}-1\right)} \; . 
\label{zlimit}
\eeq

The term $\left[ f\circ D\circ\widehat{\sigma}_k\right]$ in 
Eq.~(\ref{eq3.1}) takes the following form for the processes in (\ref{eq1.1}):

\vspace{0.5cm}

\noindent
(i) $e+p \rightarrow e + \pi + X$~\cite{Mende78}:\\
\beq
\hspace*{-6mm}\left[f\circ D\circ \widehat{\sigma}_k\right]
&=&\sum_{q,\bar{q}}e_q^2  f_q(x) D_q(z)\widehat{\sigma}^k_{qq}
+\sum_{q,\bar{q}}e_q^2  f_g(x) D_q(z)\widehat{\sigma}^k_{qg}+
\sum_{q,\bar{q}}e_q^2  f_q(x) D_g(z)\widehat{\sigma}^k_{gq} \; ,
\label{eq3.average}
\eeq
where ($C_F=4/3$)
\beq
\widehat{\sigma}^1_{qq}&=&2C_F\xhat\zhat\left\{{1\over Q^2q_T^2}
\left({Q^4\over \xhat^2\zhat^2} + \left(Q^2-q_T^2\right)^2\right) +6\right\} \; ,
\nonumber\\
\widehat{\sigma}^2_{qq}&=&2\widehat{\sigma}_4^{qq}=8C_F\xhat\zhat \; ,\nonumber\\
\widehat{\sigma}^3_{qq}&=&4C_F\xhat\zhat{1\over Qq_T}(Q^2+q_T^2) \; ,
\label{eq3.avqq}
\eeq
\beq
\widehat{\sigma}^1_{qg}&=&\xhat(1-\xhat)\left\{{Q^2\over q_T^2}
\left({1\over \xhat^2\zhat^2} -{2\over \xhat\zhat}
+2\right) +10 -{2\over \xhat}-{2\over \zhat}\right\}\; ,
\nonumber\\
\widehat{\sigma}^2_{qg}&=&2\widehat{\sigma}^4_{qg}=8\xhat(1-\xhat) \; ,\nonumber\\
\widehat{\sigma}^3_{qg}&=&\xhat(1-\xhat){2\over Qq_T}\left\{2(Q^2+q_T^2)
-{Q^2\over \xhat\zhat}\right\} \; ,
\label{eq3.avgq}
\eeq
\beq
\widehat{\sigma}^1_{gq}&=&2C_F\xhat(1-\zhat)\left\{{1\over Q^2q_T^2}
\left({Q^4\over \xhat^2\zhat^2} + {(1-\zhat)^2\over \zhat^2}
\left(Q^2-{\zhat^2q_T^2\over (1-\zhat)^2}
\right)^2\right) +6\right\} \; ,
\nonumber\\
\widehat{\sigma}^2_{gq}&=&
2\widehat{\sigma}^4_{gq}=8C_F\xhat(1-\zhat) \; ,\nonumber\\
\widehat{\sigma}^3_{gq}&=&-4C_F\xhat(1-\zhat)^2{1\over \zhat Qq_T}
\left\{Q^2+
{\zhat^2 q_T^2\over (1-\zhat)^2}\right\} \; .
\label{eq3.avqg}
\eeq

\noindent
(ii) $e+\vec{p} \rightarrow e + \vec{\Lambda} + X$:\\
\beq
\left[f\circ D\circ \widehat{\sigma}_k\right]
&=&\sum_{q,\bar{q}}e_q^2 \Delta  f_q(x)\Delta D_q(z)\Delta_L\widehat{\sigma}^k_{qq}
+\sum_{q,\bar{q}}e_q^2 \Delta  f_g(x)\Delta D_q(z)\Delta_L\widehat{\sigma}^k_{qg}
\nonumber\\
& &\qquad\qquad+\sum_{q,\bar{q}}e_q^2 \Delta  f_q(x)\Delta D_g(z)
\Delta_L\widehat{\sigma}^k_{gq} \; ,
\eeq
where
\beq
\Delta_L\widehat{\sigma}^k_{qq}=\widehat{\sigma}^k_{qq}\qquad(k=1,2,3,4) \; ,
\eeq
\beq
\Delta_L\widehat{\sigma}^1_{qg}&=&
-{(2\xhat-1)\left\{ Q^4 (\xhat-1)^2-q_T^4\xhat^2\right\}\over
Q^2 q_T^2 \xhat(\xhat-1)} \; ,
\nonumber\\
\Delta_L\widehat{\sigma}^2_{qg}&=&\Delta_L\widehat{\sigma}_4^{gq}=0 \; ,\nonumber\\
\Delta_L\widehat{\sigma}^3_{qg}&=&-{2\left\{Q^2(\xhat-1)-q_T^2\xhat\right\}
\over Qq_T} \; ,
\eeq
\beq
\Delta_L\widehat{\sigma}^1_{gq}&=&2C_F\xhat\zhat\left\{
{\xhat-2\over \xhat-1} + {\xhat(\xhat+1)\over (\xhat-1)^2}
{q_T^4\over Q^4} +{2(2\xhat^2-2\xhat +1)\over (\xhat-1)^2}{q_T^2\over Q^2}
\right\} \; ,\nonumber\\
\Delta_L\widehat{\sigma}^2_{gq}&=&2\Delta_L\widehat{\sigma}_4^{qg}=
8C_F{\xhat^2\zhat\over \xhat-1}{q_T^2\over Q^2} \; ,\nonumber\\
\Delta_L\widehat{\sigma}^3_{gq}&=&{4C_F\xhat\zhat\over (\xhat-1)^2}
\left\{(\xhat-1)^2 + {\xhat^2q_T^2\over Q^2}\right\}{q_T\over Q} \; .
\eeq

\noindent
(iii) $e+{p}^\uparrow \rightarrow e + {\Lambda}^\uparrow + X$:\\
For this process, it is more transparent to write the cross section
by including the factors ${\cal A}_k$ in (\ref{eq3.1}):
\beq
\sum_{k=1,\cdots,4,8,9}{\cal A}_k
\left[f\circ D\circ \widehat{\sigma}_k\right]
\equiv \sum_{q,\bar{q}}
e_q^2 \delta  f_q(x)\delta D_q(z)\delta\widehat{\sigma}^{qq} \; ,
\eeq
where
\beq
\delta\widehat{\sigma}^{qq} &=&
4C_F\left[ \left(1+\cosh^2\psi\right)\cos(\Phi_A-\Phi_B) - {Q\over q_T}
\sinh 2\psi\,\cos(\Phi_A-\Phi_B-\phi)\right.\nonumber\\
& &\left.+ {Q^2\over q_T^2}\sinh^2\psi\,\cos(\Phi_A-\Phi_B-2\phi)\right] \; .
\label{eq_TT}
\eeq
Here the terms with 
${\cal A}_{3}$ and ${\cal A}_{8}$ in Eq.~(\ref{eq3.1})
have been combined to give the 
$\cos(\Phi_A-\Phi_B-\phi)$ term in (\ref{eq_TT}).  Likewise,
those with ${\cal A}_{4,9}$ give the term with $\cos(\Phi_A-\Phi_B-2\phi)$.  

\noindent
(iv) $\vec{e}+\vec{p} \rightarrow e + \pi + X$:\\
\beq
\left[f\circ D\circ \widehat{\sigma}_k\right]
&=&\sum_{q,\bar{q}}e_q^2 \Delta  f_q(x) D_q(z)\Delta_{LO}\widehat{\sigma}^k_{qq}
+\sum_{q,\bar{q}}e_q^2 \Delta  f_g(x) D_q(z)\Delta_{LO}\widehat{\sigma}^k_{qg}
\nonumber\\
& &\qquad\qquad
+\sum_{q,\bar{q}}e_q^2 \Delta  f_q(x) D_g(z)\Delta_{LO}\widehat{\sigma}^k_{gq} \; ,
\eeq
where
\beq
\Delta_{LO}\widehat{\sigma}^6_{qq}&=&-2C_F\left\{
\left({1\over \xhat\zhat} +\xhat\zhat\right){Q^2\over q_T^2}
-{\xhat\zhat q_T^2\over Q^2}\right\} \; ,\nonumber\\
\Delta_{LO}\widehat{\sigma}^7_{qq}&=&-4C_F\xhat\zhat{Q^2-q_T^2\over Q q_T} \; ,
\eeq
\beq
\Delta_{LO}\widehat{\sigma}^6_{qg}&=&{2\xhat-1\over \xhat}
\left( 2\xhat + {\xhat-1\over \zhat^2}{Q^2\over q_T^2}\right) \; ,\nonumber\\
\Delta_{LO}\widehat{\sigma}^7_{qg}&=&{2Q\over q_T}{
(\xhat-1)(2\zhat -1)\over \zhat} \; ,
\eeq
\beq
\Delta_{LO}\widehat{\sigma}^6_{gq}&=&{2C_F\zhat\over \xhat-1}\left\{
{1\over \zhat^2} - (\xhat-1)^2 + {\xhat^4\over(\xhat-1)^2}
{q_T^4\over Q^4}\right\} \; ,\nonumber\\
\Delta_{LO}\widehat{\sigma}^7_{gq}&=&{4C_F\xhat\zhat\over \xhat-1}
\left(1-{\xhat\over \zhat}\right){q_T\over Q} \; .
\eeq

\noindent
(v) $\vec{e}+{p} \rightarrow e + \vec{\Lambda} + X$:\\
\beq
\left[f\circ D\circ \widehat{\sigma}_k\right]
&=&\sum_{q,\bar{q}}e_q^2  f_q(x)\Delta D_q(z)\Delta_{OL}\widehat{\sigma}^k_{qq}
+\sum_{q,\bar{q}}e_q^2  f_g(x)\Delta D_q(z)\Delta_{OL}\widehat{\sigma}^k_{qg}
\nonumber\\
& &\qquad\qquad
+\sum_{q,\bar{q}}e_q^2  f_q(x)\Delta D_g(z)\Delta_{OL}\widehat{\sigma}^k_{gq} \; ,
\eeq
where
\beq
\Delta_{OL}\widehat{\sigma}^{6,7}_{qq}=\Delta_{LO}\widehat{\sigma}^{6,7}_{qq} \; ,
\eeq
\beq
\Delta_{OL}\widehat{\sigma}^6_{qg}&=&{2\xhat^2-2\xhat+1\over \xhat\zhat}
\left( \xhat + (\xhat-1){Q^2\over q_T^2}\right) \; ,\nonumber\\
\Delta_{OL}\widehat{\sigma}^7_{qg}&=&{2Q\over q_T}{
(\xhat-1)(2\xhat -1)\over \zhat} \; ,
\eeq
\beq
\Delta_{OL}\widehat{\sigma}^6_{gq}&=&{2C_F\zhat\over \xhat-1}\left\{
{1\over \zhat^2} + (\xhat-1)^2 - {\xhat^4\over(\xhat-1)^2}
{q_T^4\over Q^4}\right\} \; ,\nonumber\\
\Delta_{OL}\widehat{\sigma}^7_{gq}&=&-{4C_F\xhat\zhat\over \xhat-1}
\left(1-{\xhat\over \zhat}\right){q_T\over Q} \; .
\eeq

\end{document}